# FINDING LARGE AVERAGE SUBMATRICES IN HIGH DIMENSIONAL DATA

By Andrey A. Shabalin,[1] Victor J. Weigman,
Charles M. Perou[2] and Andrew B. Nobel

*University of North Carolina at Chapel Hill*

The search for sample-variable associations is an important problem in the exploratory analysis of high dimensional data. Biclustering methods search for sample-variable associations in the form of distinguished submatrices of the data matrix. (The rows and columns of a submatrix need not be contiguous.) In this paper we propose and evaluate a statistically motivated biclustering procedure (LAS) that finds large average submatrices within a given real-valued data matrix. The procedure operates in an iterative-residual fashion, and is driven by a Bonferroni-based significance score that effectively trades off between submatrix size and average value. We examine the performance and potential utility of LAS, and compare it with a number of existing methods, through an extensive three-part validation study using two gene expression datasets. The validation study examines quantitative properties of biclusters, biological and clinical assessments using auxiliary information, and classification of disease subtypes using bicluster membership. In addition, we carry out a simulation study to assess the effectiveness and noise sensitivity of the LAS search procedure. These results suggest that LAS is an effective exploratory tool for the discovery of biologically relevant structures in high dimensional data.

Software is available at https://genome.unc.edu/las/.

**1. Introduction.** Unsupervised exploratory analysis plays an important role in the study of large, high-dimensional datasets that arise in a variety of applications, including gene expression microarrays. Broadly speaking, the

Received July 2008; revised December 2008.

[1]Supported in part by National Institutes of Health Grant P42 ES005948 and R01 AA016258.

[2]Supported in part by funds from the NCI Breast SPORE program to UNC-CH (P50-CA58223-09A1), by the V Foundation for Cancer Research, and by the Breast Cancer Research Foundation.

goal of such analysis is to find patterns or regularities in the data, without *ab initio* reference to external information about the available samples and variables. One important source of regularity in experimental data are associations between sets of samples and sets of variables. These associations correspond to distinguished submatrices of the data matrix, and are generally referred to as biclusters, or subspace clusters. In gene expression and related analyses, biclusters, in conjunction with auxiliary clinical and biological information, can provide a first step in the process of identifying disease subtypes and gene regulatory networks.

In this paper we propose and evaluate a statistically motivated biclustering procedure that finds large average submatrices within a given real-valued data matrix. The procedure, which is called LAS (for Large Average Submatrix), operates in an iterative fashion, and is based on a simple significance score that trades off between the size of a submatrix and its average value. A connection is established between maximization of the significance score and the minimum description length principle.

We examine the performance and utility of LAS, and compare it with a number of existing methods, through an extensive validation study using two independent gene expression datasets. The validation study has three parts. The first concerns quantitative properties of the biclustering methods such as bicluster size, overlap and coordinate-wise statistics. The second is focused biological and clinical assessments using auxiliary information about the samples and genes under study. In the the third part of the study, the biclusters are used to perform classification of disease subtypes based in their sample membership. In addition, we carry out a simulation study to assess the effectiveness and noise sensitivity of the LAS search procedure.

1.1. *Biclustering.* Sample-variable associations can be defined in a variety of ways, and can take a variety of forms. The simplest, and most common, way of identifying associations in gene expression data is to independently cluster the rows and columns of the data matrix using a multivariate clustering procedure [Weinstein et al. (1997); Eisen et al. (1998); Tamayo et al. (1999); Hastie et al. (2000)]. When the rows and columns of the data matrix are reordered so that each cluster forms a contiguous group, the result is a partition of the data matrix into nonoverlapping rectangular cells. The search for sample variable associations then consists of identifying cells whose entries are, on average, bright red (large and positive) or bright green (large and negative) [Weigelt et al. (2005)]. In some cases, one can improve the results of independent row–column clustering by simultaneously clustering samples and variables, a procedure known as co-clustering [Hartigan (1972); Kluger et al. (2003); Dhillon (2001); Getz, Levine and Domany (2000)].

Independent row–column clustering (IRCC) has become a standard tool for the visualization and exploratory analysis of microarray data, but it is an indirect approach to the problem of finding sample-variable associations. By contrast, biclustering methods search directly for sample-variable associations, or more precisely, for submatrices $U$ of the data matrix $X$ whose entries meet a predefined criterion. Submatrices meeting the criterion are typically referred to as biclusters. It is important to note that the rows and columns of a bicluster (and more generally a submatrix) need not be contiguous. A number of criteria for defining biclusters $U$ have been considered in the literature, for example: the rows of $U$ are approximately equal to each other [Aggarwal et al. (1999)]; the columns of $U$ are approximately equal [Friedman and Meulman (2004)]; the elements of $U$ are well-fit by a 2-way ANOVA model [Cheng and Church (2000); Lazzeroni and Owen (2002); Wang et al. (2002)]; the rows of $U$ have equal [Ben-Dor et al. (2003)] or approximately equal [Liu, Yang and Wang (2004)] rank statistics; and all elements of $U$ are above a given threshold [Prelic et al. (2006)].

The focus of this paper is the simple criterion that the average of the entries of the submatrix $U$ is large and positive, or large and negative. Submatrices of this sort will appear red or green in the standard heat map representation of the data matrix, and are similar to those targeted by independent row–column clustering.

1.2. *Features of biclustering.* While its direct focus on finding sample-variable associations makes biclustering an attractive alternative to row–column clustering, biclustering has a number of other features, both positive and negative, that we briefly discuss below.

Row-column clustering assigns each sample and each variable to a unique cluster. By contrast, the submatrices produced by biclustering methods may overlap, and need not cover the entire data matrix, features that better reflect the structure of many scientific problems. For example, the same gene can play a role in multiple pathways, and a single sample may belong to multiple phenotypic or genotypic subtypes. Multiple bicluster membership for rows and columns can directly capture this aspect of experimental data.

In row–column clustering, the group to which a sample is assigned depends on all the available variables, and the group to which a variable is assigned depends on the all the available samples. By contrast, biclusters are locally defined: the inclusion of samples and variables in a bicluster depends only on their expression values inside the associated submatrix. Locality allows biclustering methods to target relevant genes and samples while ignoring others, giving such methods greater exploratory power and flexibility than row–column clustering. For more on the potential advantages of biclustering, see Madeira and Oliveira (2004); Jiang, Tang and Zhang (2004); Parsons, Haque and Liu (2004).

Figure 1 illustrates the differences between the blocks arising from independent row–column clustering and those arising from biclustering. Note that while one may display an individual bicluster as a contiguous block of variables and samples by suitably reordering the rows and columns of the data matrix, when considering more than two biclusters, it is not always possible to display them simultaneously as contiguous blocks.

The flexibility and exploratory power of biclustering methods comes at the cost of increased computational complexity. Most biclustering problems are NP complete, and even the most efficient exact algorithms (those that search for every maximal submatrix satisfying a given criterion) can be prohibitively slow, and produce a large number of biclusters, when they are applied to large datasets. The LAS algorithm relies on a heuristic (nonexact), randomized search to find biclusters, a feature shared by many existing methods.

**2. The LAS algorithm.** In this paper we present and assess a significance-based approach to biclustering of real-valued data. Using a simple Gaussian null model for the observed data, we assign a significance score to each submatrix $U$ of the data matrix using a Bonferroni-corrected $p$-value that is based on the size and average value of the entries of $U$. The Bonferroni correction accounts for multiple comparisons that arise when searching among many submatrices for a submatrix having a large average value. In addition, the correction acts as a penalty that controls the size of discovered submatrices. (Connections between LAS and the Minimum Description Length principle are discussed in Section 2.3 below.)

2.1. *Basic model and score function.* Let $X = \{x_{i,j} : i \in [m], j \in [n]\}$ be the observed data matrix. (Here and in what follows, $[k]$ denotes the set of integers from 1 to $k$.) A submatrix of $X$ is an indexed set of entries $U = \{x_{i,j} : i \in A, j \in B\}$ associated with a specified set of rows $A \subseteq [m]$ and columns $B \subseteq [n]$. In general, the rows in $A$ and the columns in $B$ need not be contiguous.

The LAS algorithm is motivated by an additive submatrix model under which the data matrix $X$ is expressed as the sum of $K$ constant, and potentially overlapping, submatrices plus noise. More precisely, the model states

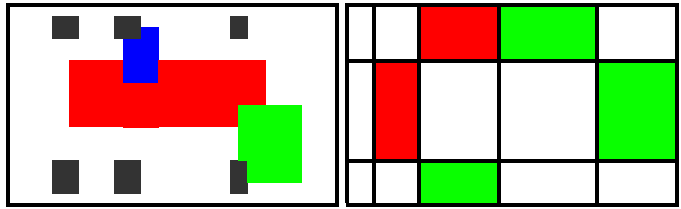

Fig. 1. *Illustration of bicluster overlap (left) and row–column clustering (right).*

that

(1) $$x_{i,j} = \sum_{k=1}^{K} \alpha_k I(i \in A_k, j \in B_k) + \varepsilon_{ij}, \qquad i \in [m], j \in [n],$$

where $A_k \subseteq [m]$ and $B_k \subseteq [n]$ are the row and column sets of the $k$th submatrix, $\alpha_k \in \mathbb{R}$ is the level of the $k$th submatrix, and $\{\varepsilon_{ij}\}$ are independent $N(0,1)$ random variables. Here $I(\cdot)$ is an indicator function equal to one when the condition in parentheses holds. When $K = 0$, the model (1) reduces to the simple null model

(2) $$\{x_{i,j} : i \in [m], j \in [n]\} \text{ are i.i.d.} \qquad \text{with } x_{i,j} \sim \mathcal{N}(0,1)$$

under which $X$ is an $m \times n$ Gaussian random matrix.

The null model (2) leads naturally to a significance based score function for submatrices. In particular, the score assigned to a $k \times l$ submatrix $U$ of $X$ with average $\mathrm{Avg}(U) = \tau > 0$ is defined by

(3) $$S(U) = -\log\left[\binom{m}{k}\binom{n}{l}\Phi(-\tau\sqrt{kl})\right].$$

The term in square brackets is an upper bound on the probability of the event $A$ that there exists a $k \times l$ submatrix with average greater than or equal to $\tau$ in an $m \times n$ Gaussian random matrix. More precisely, by the union bound, $P(A) \leq \sum P(\mathrm{Avg}(V) \geq \tau)$, where the sum ranges over all $k \times l$ submatrices $V$ of $X$. Under the Gaussian null, each probability in the sum is equal to $\Phi(-\tau\sqrt{kl})$, so that $P(A) \leq N\Phi(-\tau\sqrt{kl})$, where $N = \binom{m}{k}\binom{n}{l}$ is the number of $k \times l$ submatrices of an $m \times n$ matrix. From a testing point of view, the term in brackets can be thought of as a Bonferroni corrected $p$-value associated with the null model (2) and the test function $\mathrm{Avg}(U)$.

The score function $S(\cdot)$ measures departures from the null (2) in a manner that accounts for the dimensions and average value of a submatrix. It provides a simple, one-dimensional yardstick with which one can compare and rank observed submatrices of different sizes and intensities. Among submatrices of the same dimensions, it favors those with higher average.

2.2. *Description of algorithm.* The LAS score function is based on the normal CDF, and is sensitive to departures from normality that arise from heavy tails in the empirical distribution of the expression values. Outliers can give rise to submatrices that, while highly significant, have very few samples or variables. As a first step in the algorithm we consider the standard Q–Q plot of the empirical distribution of the entries of the column-standardized data matrix against the standard normal CDF. Both the breast cancer and lung cancer datasets considered in Section 4 exhibited heavy tails. To address this, we applied the transformation $f(x) = \mathrm{sign}(x)\log(1 + |x|)$ to each entry of the data matrix. After transformation, the Q–Q plot indicated excellent agreement with the normal distribution. Other alternatives to the squashing function $f(\cdot)$ can also be considered.

The LAS algorithm initially searches for a submatrix of $X$ maximizing the significance score $S(\cdot)$. Once a candidate submatrix $U^*$ has been identified, a residual matrix $X'$ is computed by subtracting the average of $U^*$ from each of its elements in $X$. Formally, $x'_{i,j} = x_{i,j} - \text{Avg}(U^*)$ if $x_{i,j}$ is in $U^*$, and $x'_{i,j} = x_{i,j}$ otherwise. The search procedure is then repeated on the residual matrix $X'$. The core of the algorithm is a randomly initialized, iterative search procedure for finding a maximally significant submatrix of a given matrix. The pseudo code for the algorithm is as follows:

*Input*: Data matrix $X$.
*Search*: Find a submatrix $U^*$ of $X$ that approximately maximizes the score function $S(\cdot)$.
*Residual*: Subtract the average of $U^*$ from each of its elements in $X$.
*Repeat*: Return to Search.
*Stop*: When $S(U^*)$ falls below a threshold, or a user-defined number of submatrices are produced.

The output of the algorithm is a collection of submatrices having significant positive averages. Repeating the algorithm with $X$ replaced by $-X$ yields submatrices with significant negative averages.

It is not feasible in the search procedure to check the score of each of the $2^{n+m}$ possible submatrices of $X$. Instead, the procedure iteratively updates the row and column sets of a candidate submatrix in a greedy fashion until a local maximum of the score function is achieved. For fixed $k, l$, the basic search procedure operates as follows:

*Initialize*: Select $l$ columns of $B$ at random.
*Loop*: Iterate until convergence of $A$, $B$:
  Let $A := k$ rows with the largest sum over the columns in $B$.
  Let $B := l$ columns with the largest sum over the rows in $A$.
*Output*: Submatrix associated with final $A$, $B$.

As currently implemented, the initial values of $k$ and $l$ are selected at random from the sets $\{1, \ldots, \lceil m/2 \rceil\}$ and $\{1, \ldots, \lceil n/2 \rceil\}$ respectively, and are held fixed until the algorithm finds a local maximum of the score function. On subsequent iterations, the algorithm adaptively selects the number of rows and columns in order to maximize the significance score. Each run of the basic search procedure yields a submatrix that is a local maximum of the score function, *that is*, a submatrix that cannot be improved by changing only its column set or its row set. The basic search procedure is repeated 1000 times, and the most significant submatrix found is returned in the main loop of the algorithm. In experiments on real data (see Section 5.3), we found that 1000 iterations of the main loop of the search procedure was sufficient to ensure stable performance of the algorithm.

Many biclustering methods require the user to specify a number of operational parameters, and in many cases, getting optimal performance from the method can require careful choice and tuning of the parameters. In addition, for exact algorithms, minor alteration of the parameters can result in substantial changes in the size and interpretability of the output. The only operational parameters of the LAS algorithm are the number of times the basic search procedure is run in each main loop of the algorithm, and the stopping criterion. This minimal number of parameters is an important feature of LAS, one that makes application of the method to scientific problems relatively straightforward.

2.3. *Penalization and MDL.* The score function employed by LAS can be written as a sum of two terms. The first, $-\log \Phi(-\sqrt{kl}\tau)$, is positive and can be viewed as a "reward" for finding a $k \times l$ submatrix with average $\tau$. The second, $-\log[\binom{m}{k}\binom{n}{l}]$, is negative and is a multiple comparisons penalty based on the number of $k \times l$ submatrices in $X$. The penalty depends separately on $k$ and $l$, and its combinatorial form suggests a connection with the Minimum Description Length Principle (MDL), following Rissanen (2004), Grunwald (2004), and Barron and Yu (1998). The MDL principle is a formalization of Occam's Razor, in which the best model for a given set of data is the one that leads to the shortest overall description of the data.

In the Appendix we describe a code for matrices based on a family of additive models, and show that the description length of a matrix with an elevated submatrix is approximately equal to a linear function of its LAS score. The penalty term in the LAS score function corresponds to the length of the code required to describe the location of a $k \times l$ submatrix, while the "reward" is related to the reduction in code length achieved by describing the residual matrix instead of the original matrix. The connection with MDL provides support for the significance based approach to biclustering adopted here.

**3. Description of competing methods.** In this section we describe the methods to which we will compare the LAS algorithm in the validation sections below. We considered biclustering methods that search directly for sample variable associations, as well as biclusters derived from independent row–column clustering.

3.1. *Biclustering methods.* Initially, we compared LAS with six existing biclustering methods: Plaid, CC, SAMBA, ISA, OPSM, and BiMax. These methods employ a variety of objective functions and search algorithms. We limited our comparisons to methods that (i) have publicly available implementations with straightforward user interfaces, (ii) can efficiently handle large datasets arising from gene expression and metabolomic data, and (iii) are well suited to use by biologists. The methods are described in more detail below.

The Plaid algorithm of Lazzeroni and Owen (2002) employs an iterative procedure to approximate the data matrix $X$ by a sum of submatrices whose entries follow a two-way ANOVA model. At each stage, Plaid searches for a submatrix maximizing explained variation, as measured by reduction in the overall sum of squares. We set the parameters of Plaid so that at each stage it fits a constant submatrix (with no row or column effects). With these settings, the Plaid method is most closely related to LAS, and also derives from a block-additive model like (1). We have also run Plaid with settings under which it fits biclusters by a general ANOVA model. The two versions of Plaid exhibit similar validation results; we present only those for which Plaid fits biclusters by a constant. Various modifications of the Plaid model and algorithm have been proposed in the literature: Turner, Bailey and Krzanowski (2005) have developed an improved algorithm and Segal, Battle and Koller (2003), Gu and Liu (2008), and Caldas and Kaski (2008) have considered the Plaid problem in the Bayesian framework. We have chosen to focus on the original Plaid algorithm of Lazzeroni and Owen (2002), as it is both the first and most representative method of its type.

The Cheng and Church (CC) biclustering algorithm [Cheng and Church (2000)] searches for submatrices such that the sum of squared residuals from a two-way ANOVA fit falls below a given threshold. These biclusters are locally maximal in the sense that addition of any more rows or columns will increase the mean squared error beyond the threshold. Whereas Plaid searches for a submatrix maximizing explained variation, CC searches for large submatrices with small unexplained variation. The LAS, Plaid, and CC algorithms discover biclusters sequentially. Once a candidate target is identified, LAS and Plaid form the associated residual matrix before proceeding to the next stage. By contrast, CC replaces the values of the target submatrix by Gaussian noise.

The SAMBA algorithm of Tanay, Sharan and Shamir (2002) adopts a graph theoretic approach, in which the data matrix is organized into a bipartite graph, with one set of nodes corresponding to genes and the other corresponding to samples. Weights are then assigned to edges that connect genes and samples based on the data matrix, and the subgraphs with the largest overall weights are returned.

Ihmels et al. (2002) developed a biclustering algorithm (ISA) that searches for maximal submatrices whose row and column averages exceed preset thresholds. Both LAS and ISA rely on iterative search procedures that are variants of EM and Gibbs type algorithms. In both methods, the search procedure alternately updates the columnset (given the current rowset) and then the rowset (given the current columnset) until converging to a local optimum.

The OPSM algorithm Ben-Dor et al. (2003) searches for maximal submatrices whose rows have the same order statistics. Like LAS, the OPSM algorithm makes use of a multiple comparison corrected $p$-value in assessing and comparing biclusters of different sizes.

Each of the algorithms above employs heuristic strategies to search for distinguished submatrices. By contrast, the Bimax algorithm of Prelic et al. (2006) uses a divide-and-conquer approach to find *all* inclusion-maximal biclusters whose values are above a user-defined threshold. Bimax is the only exact algorithm among those considered here.

We ran all methods except Plaid and CC with their default parameter settings. LAS, CC, and Plaid allow the user to choose the number of biclusters produced; we selected 60 biclusters for each method. The settings of Plaid were chosen so that the submatrix fit at each stage is a constant, without row and column effects. Once the CC method identifies a bicluster, it removes it from the data matrix by replacing its elements by noise. When the CC method was run with the default parameter $\delta = 0.5$, it initially produced a single bicluster that contained most of the available genes and samples, leaving very little information from which additional biclusters could be identified. To solve this problem, we reduced the $\delta$ parameter in CC from 0.5 to 0.1. A description of the computer used to run the biclustering algorithms and the set of parameters used for each method is provided in the Appendix.

3.2. *Independent row–column clustering (IRCC).* In addition to the methods described above, we also produced biclusters from $k$-means and hierarchical clustering. We applied $k$-means clustering independently to the rows and columns of the data matrix, with values of $k$ ranging from 3 to 15. In each case, we produced 30 clusterings and selected the one with the lowest sum of within-cluster sum of squares. The set of $85 \times 117 = 9945$ submatrices (not all column clusters were unique) obtained from the Cartesian product of the row and column clusters is denoted by KM.

We applied hierarchical clustering independently to the rows and columns of the data matrix using a Pearson correlation based distance and average linkage. All clusters associated with subtrees of the dendrogram were considered, but row clusters with less that 10 rows, and column clusters with less than 8 columns, were discarded. The resulting set of $34 \times 2806 = 95{,}404$ submatrices obtained from the Cartesian product of the row and column clusters is denoted by HC.

**4. Comparison and validation.** Existing biclustering methods differ widely in their underlying criteria, as well as the algorithms they employ to identify biclusters that satisfy these criteria. As such, simulations based on the additive submatrix model (1) cannot fairly be used to assess the performance of competing methods that are based on different models and submatrix criteria. For this reason our assessment of LAS relies more heavily on biological validation rather than simulations: the former provides a direct comparison of the methods in terms of their practical utility.

We applied LAS and the biclustering methods described in the previous section to two existing gene expression datasets: a breast cancer study from Hu et al. (2006), and a lung cancer study from Bhattacharjee et al. (2001). The datasets can be downloaded from the University of North Carolina Microarray Database (UMD, http://genome.unc.edu) and http://www.broad.mit.edu/mpr/lung/ respectively. In this section we describe and implement a number of validation measures for assessing and comparing the performance of the biclustering methods under study. The validation results for the breast cancer study are detailed below; the results for the lung cancer data are contained in the Supplementary Materials [Shabalin et al. (2009)]. The validation measures considered here are applicable to any biclustering method and most gene expression type datasets.

4.1. *Description of the Hu data.* This dataset is from a previously published breast cancer study [Hu et al. (2006)] that was based on 146 Agilent 1Av2 microarrays. Initial filtering and normalization followed the protocol in Hu et al.: genes with intensity less than 30 in the red or green channel were removed; for the remaining genes, red and green channels were combined using the $\log_2$ ratio. The initial log-transformed dataset was row median centered, and missing values were imputed using a $k$-nearest neighbor algorithm with $k = 10$. Among the 146 samples, there were 29 pairs of biological replicates in which RNA was prepared from different sections of the same tumor. To avoid giving these samples more weight in the analysis, we removed the replicates, keeping only the primary tumor profiles. After preprocessing, the dataset contained 117 samples and 13,666 genes. In what follows, the dataset will be referred to as *Hu*.

4.2. *Quantitative comparisons.* LAS, Plaid, and CC were set to produce 60 biclusters. The number of biclusters produced by other methods was determined by their default parameters, with values ranging from 15 (OPSM) to 1977 (BiMax). KM and HC produced 9945 and 95,404 biclusters, respectively. Table 1 shows the number of biclusters produced by each method.

All biclustering methods were run on the same computer, having an AMD64 FX2 Dual Core processor with 4GB of RAM (a complete specification is provided in the Appendix). The running time of LAS was 85 minutes;

TABLE 1
*Output summary for different biclustering methods. From left to right: total number of biclusters produced; effective number of biclusters; the ratio of the effective number to the total number of biclusters*

| Method | # of clusters | Eff. # of clusters | Ratio |
|---|---|---|---|
| LAS | 60 | 48.6 | 0.810 |
| Plaid | 60 | 6.4 | 0.106 |
| CC | 60 | 60.0 | 1.000 |
| ISA | 72 | 42.3 | 0.588 |
| OPSM | 15 | 9.1 | 0.605 |
| SAMBA | 289 | 171.7 | 0.594 |
| BiMax | 1977 | 42.9 | 0.022 |
| KM | 9945 | 78.7 | 0.008 |
| HC | 95,404 | 800.4 | 0.008 |

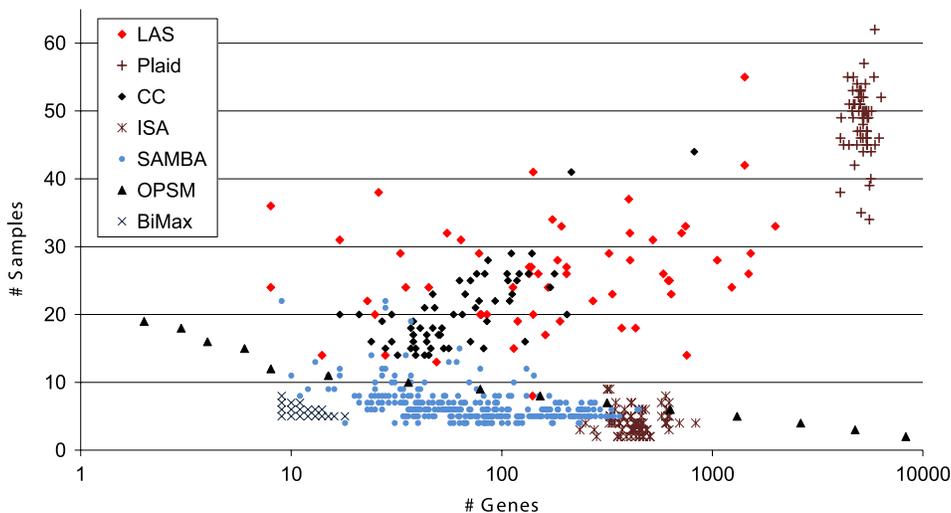

FIG. 2. *Bicluster sizes for different methods.*

ISA and OPSM finished in about 30 minutes; CC, Plaid, and SAMBA finished in less then 10 minutes. The Bimax algorithm took approximately 5 days. Hierarchical clustering took 2 minutes, while $k$-means clustering (with $k = 3, \ldots, 15$ and 30 repeats) took 1 hour 40 minutes. Our primary focus in validation was output quality.

4.2.1. *Bicluster sizes.* In Figure 2 we plot the row and column dimensions of the biclusters produced by the different methods. The resulting scatter plot shows marked differences between the methods under study, and provides useful insights into their utility and potential biological findings. (A similar figure could be used, for example, to assess the effects of different parameter settings for a single method of interest.) Both LAS and CC produce a relatively wide range of bicluster sizes, with those of LAS ranging from $8 \times 8$ (genes $\times$ samples) to $1991 \times 55$. The other methods tested produced biclusters with a more limited range of sizes. Biclusters produced by ISA, OPSM, and SAMBA have a relatively small number of samples, less than 10 samples per bicluster on average in each case. (Some of the points denoting OPSM clusters have been obscured in the figure.) The biclusters produced by Bimax had at most 8 samples, and at most 18 genes. By contrast, Plaid produced large biclusters, having an average of 49 samples and 5130 genes per bicluster.

The differences between LAS and Plaid bear further discussion. We ran Plaid with settings (constant fit, no row, and column effects) that made it most similar to LAS. With these settings, both methods rely on similar models, and proceed in stages via residuals, but differ in their objective functions. Plaid seeks to maximize the explained variation $kl\tau^2$, or equivalently, $-\log \Phi(-\sqrt{kl}\tau)$. By contrast, the score function maximized by LAS includes a combinatorial penalty term involving $k$ and $l$ that acts to control the size of the discovered submatrices. In this, and other, experiments, the penalty excludes very large submatrices, and produces a relatively wide range of bicluster sizes. (While the combinatorial penalty is small for values of $k$ close to $m$ and $l$ close to $n$, submatrices of this size tend to have a small average value.)

4.2.2. *Effective number of biclusters.* Distinct biclusters produced by the same method may exhibit overlap. On the one hand, the flexibility of overlapping gene and sample sets has the potential to better capture underlying biology. On the other hand, extreme overlap of biclusters can reduce a method's effective output: two moderate sized biclusters that differ only in a few rows or columns do not provide much more information than either bicluster alone. Whatever the source of overlap, it is helpful to keep it in mind when evaluating other features of a method, such as the number of biclusters it produces that are deemed to be statistically significant. To this end, we measure the effective number of biclusters in a family $U_1, \ldots, U_K$ by

$$F(U_1, \ldots, U_K) = \sum_{k=1}^{K} \frac{1}{|U_k|} \sum_{x \in U_k} \frac{1}{N(x)},$$

where $N(x) = \sum_{k=1}^{K} I\{x \in U_k\}$ is the number of biclusters containing matrix entry $x$. The measure $F(\cdot)$ has the property that if, for any $1 \leq r \leq K$, the biclusters $U_1, \ldots, U_K$ can be divided into $r$ nonoverlapping groups of identical biclusters, then $F(U_1, \ldots, U_K) = r$.

Table 1 shows the effective number of biclusters produced by each method. The low overlap of the CC algorithm is due to the fact that it replaces the values in discovered submatrices by Gaussian noise, so that a matrix element is unlikely to belong to more than one bicluster. Bimax is an exhaustive method with no pre-filtering of its output; it produced a large number of small, highly overlapping biclusters. Biclusters produced by LAS had modest levels of overlap, less than those of all other methods, except CC. The high overlap of Plaid biclusters is explained in part by their large size.

4.2.3. *Score-based comparison of LAS and standard clustering.* Ideally, a direct search for large average submatrices should improve on the results of independent row–column clustering. To test this, we computed the significance score $S(C)$ for every cluster produced by KM and HC, and compared these to the scores obtained with LAS. The highest scores achieved by KM and HC biclusters were 6316 and 5228, respectively. The first LAS biclusters had scores 12,883 (positive average) and 10,070 (negative average); the scores of the first 6 LAS biclusters were higher than scores of all the biclusters produced by KM or HC. The highest score achieved by a Plaid bicluster was 12,542, which also dominated the scores achieved by KM and HC. These results show that LAS is capable, *in practice*, of finding submatrices that cannot be identified by standard clustering methods. We also note that LAS produces only 60 biclusters, while KM and HC produce 9945 and 95,404 biclusters, respectively.

4.2.4. *Summary properties of row and column sets.* One potential benefit of biclustering methods over independent row–column clustering is that the sample-variable associations they identify are defined locally: they can, in principle, identify patterns of association that are not readily apparent from the summary statistics across rows and columns that drive $k$-means and hierarchical clustering. Nevertheless, local associations can sometimes be revealed by summary measures of variance and correlation, and it is worthwhile to consider the value of these quantities as a way of seeing (a) what drives different biclustering methods, and (b) the extent to which the local discoveries of these methods are reflected in more global summaries.

For each method under study, the first four columns of Table 2 show the average, across the biclusters, of the following summary statistics: (i) the average pairwise correlation of their constituent genes, (ii) the average pairwise correlation of their constituent samples, (iii) the average standard deviation of their constituent genes, and (iv) the average standard deviation of their constituent samples. Average values for the entire matrix are shown in the first row of the table. (Recall that the data matrix is column standardized, so the column standard deviations are all equal to one.) In each case, the statistics associated with the biclustering methods are higher than the average of these statistics over the entire matrix. As HC is based entirely on gene and sample correlations, we expect its correlation values to be large compared with other methods, and this is the case. The moderate values of gene correlation for KM result from the fact that we are using a relatively small numbers of gene clusters, which tend to have a large number of genes and therefore low average pairwise correlations. Similar remarks apply to the low gene (and sample) correlation values associated with Plaid.

Table 2
*Average standard deviation and average pairwise correlation of genes and samples, for biclusters, KM and HC clusters, and the whole data matrix. As a reference point, the last row shows the summary statistics for samples belonging to the same disease subtype*

|  | Correlation | | Std. Deviation | | |
|---|---|---|---|---|---|
|  | **Gene** | **Sample** | **Gene** | **Sample** | **Within variance** |
| Matrix | 0.00 | 0.01 | 0.89 | 1.00 | 1.00 |
| LAS | 0.34 | 0.10 | 1.40 | 1.00 | 1.96 |
| Plaid | 0.02 | 0.03 | 0.99 | 1.00 | 1.24 |
| CC | 0.09 | 0.05 | 1.02 | 1.00 | 0.49 |
| ISA | 0.22 | 0.31 | 0.99 | 1.00 | 1.99 |
| OPSM | 0.48 | 0.06 | 0.93 | 1.00 | 1.18 |
| SAMBA | 0.26 | 0.02 | 1.66 | 1.00 | 3.36 |
| BiMax | 0.09 | 0.26 | 3.42 | 1.00 | 27.75 |
| KM | 0.19 | 0.22 | 0.91 | 1.00 | 0.96 |
| HC | 0.44 | 0.24 | 0.93 | 1.00 | 0.88 |
| Subtypes |  | 0.13 |  | 1.00 |  |

BiMax appears to be driven by all summary measures, with gene correlation playing a relatively minor role, while ISA is not affected by gene standard deviation. LAS appears to be driven by a mix of gene correlation and standard deviation. The average summary statistics of LAS do not appear to be extreme, or to reflect overtly global behavior. In each column, the average for LAS is less than and greater than those of two other methods. The remaining biclustering methods appear to depend on two, or in some cases only one, of the measured summary statistics. We note that the average pairwise correlation of the samples in LAS biclusters best matches the average pairwise correlation of samples in the cancer subtypes (described in Section 4.3.1 below).

4.2.5. *Tightness of biclusters.* For each method under consideration we calculated the average of the within bicluster variances. The results are presented in the rightmost column of Table 2. BiMax and SAMBA, which operate on thresholded entries, find biclusters with high average variance. LAS, Plaid, and ISA search for biclusters with high overall or high row/column averages; they find biclusters with variance above one. Biclusters identified by the CC algorithm have the smallest average variance, as CC searches for biclusters with low unexplained variation. The IRCC methods find biclusters with average variance only slightly lower that one.

4.3. *Biological comparisons.* The previous section compares LAS with other biclustering and IRCC methods on the basis of quantitative measures that are not directly related to biological or clinical features of the data. In this section we consider several biologically motivated comparisons. In particular, we carry out a number of tests that assess the gene and sample sets of each bicluster using auxiliary clinical information and external annotation. The next subsection considers sample-based measures of subtype capture.

4.3.1. *Subtype capture.* Breast cancer encompasses several distinct diseases, or subtypes, which are characterized by unique and substantially different expression signatures. Each disease subtype has associated biological mechanisms that are connected with its pathologic phenotype and the survival profiles of patients [cf. Golub et al. (1999); Sorlie et al. (2001), Weigelt et al. (2005), Hayes et al. (2006)]. Breast cancer subtypes were initially identified using hierarchical clustering of gene expression data, and have subsequently been validated in several datasets [cf. Fan et al. (2006)] and across platforms [cf. Sorlie et al. (2003)]. They are one focal point for our biological validation.

Hu et al. (2006) assigned each sample in the dataset to one of 5 disease subtypes (Basal-like, HER2-enriched, Luminal A, Luminal B, and Normal-like) using a nearest shrunken centroid predictor [Tibshirani et al. (2002)] and a pre-defined set of 1300 intrinsic genes. The centroids for the predictor were derived from the hierarchical clustering of 300 samples chosen both for data quality and the representative features of their expression profiles. In addition, each sample in the Hu dataset was assigned via a clinical assay to one of two estrogen receptor groups, denoted ER+ and ER−, which constitute the ER status of the tumor. The ER status of tumors is closely related to their subtypes: in the Hu dataset, HER2-enriched and Basal-like samples are primarily ER-negative (74% and 94% respectively), while Normal-like, Luminal A and Luminal B are primarily ER-positive (83%, 86%, and 91% respectively) .

Here we compare the ability of biclustering methods to capture the disease subtype and ER status of the samples. In order to assess how well the set of samples associated with a bicluster captures a particular subtype, we measured the overlap between the two sample groups using a $p$-value from a standard hypergeometric test (equivalent to a one-sided Fisher's exact test). For each biclustering method, we identified the bicluster that best matched each subtype, and recorded its associated *p*-value. As a point of comparison, we include the subtype match of column clusters produced by *k*-means and hierarchical clustering. The results are shown in Figure 3.

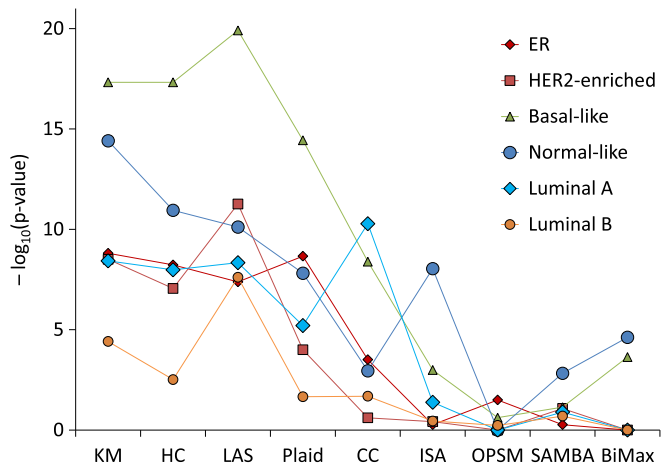

Fig. 3. *The minus $\log_{10}$ p-values of best subtype capture for different biclustering and sample clustering methods.*

The figure indicates that LAS captures ER status and disease subtypes better than the other biclustering methods, with the single exception of the Luminal A subtype, which was better captured by CC. In addition, LAS is competitive with KM and HC, performing better or as well as these methods on the Luminal A, Luminal B, Basal-like, and HER2-enriched subtypes.

Another view of subtype capture is presented in the bar-plot of Figure 4. For the Basal-like disease subtype, the figure shows the number of true, missed, and false discoveries associated with the the best sample cluster (as judged by the hypergeometric $p$-value) that was produced by each method. The Basal-like subtype contains 32 samples. The best LAS bicluster captured 27 of the 32 Basal-like samples with no false positives. Plaid had fewer missed samples, but a larger number of false positives, due to the large size of its sample clusters. As the disease subtypes were identified in part through the use of hierarchical clustering, the strong performance of KM and HC is unsurprising. Other biclustering methods were not successful in capturing Basal-like or other subtypes, due in part to the small number of samples in their biclusters. Barplots like Figure 4 for other subtypes are provided in the Supplementary Materials [Shabalin et al. (2009)].

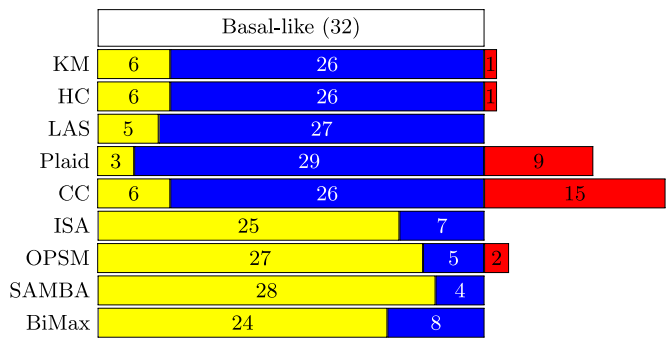

Fig. 4. *Bar-plot of missed, true, and false discoveries for different biclustering methods and the Luminal A subtype. Bars represent: light—missed discoveries, dark—true discoveries, gray—falsely discoveries.*

4.4. *Biclusters of potential biological interest.* In order to assess the potential biological and clinical relevance of the biclustering methods under consideration, we applied three different tests to the gene and sample sets of each bicluster. The first test makes use of clinical information concerning patient survival. The second tests for over-representation of functional categories and genomic neighborhoods (cytobands) among the gene sets of different biclusters, and the third tests for the differential expression of these same gene categories between the sample set of a bicluster and its complement. The tests are described in more detail below.

We chose not to include KM and HC in this analysis for several reasons. The tests conducted here are intended to provide a rough biological assessment of the gene and sample sets of biclusters that are produced with the primary goal of capturing gene-sample associations. In this sense, the tests here are assessing secondary features of these methods. By contrast, gene and sample based tests are separately assessing the primary features of KM and HC, for which biclusters are a byproduct of their independent gene and sample grouping.

For 105 samples out of 117 in the dataset, we have information regarding overall survival (OS) and relapse free survival (RFS). We applied the standard logrank test [see Bewick, Cheek and Ball (2004)] to determine if there are significant differences between the survival times associated with samples in a bicluster and the survival times associated with samples in its complement. Biclusters whose associated patients have significantly lower (or higher) survival rates are of potential clinical interest, as their gene sets may point to biological processes that play a deleterious (or beneficial) role in survival. A bicluster was called significant if its samples passed the logrank test for overall or relapse free survival at the 5% level. The number of biclusters meeting the criterion is presented in the *Survival* column of Table 3.

We next tested the gene set of each bicluster for over-representation of biologically derived functional categories and genomic neighborhoods. For the former, we considered KEGG categories [Kyoto Encyclopedia of Genes

Table 3

*The number of biclusters passing tests for survival, and gene-set enrichment and sample-set differential expression of KEGG categories and cytobands. A detailed description of the tests is given in the text*

|       | # of BC's | Survival 5% level | KEGG/Cytoband Gene | Sample | 2 out of 3 | All 3 |
|-------|-----------|-------------------|------|--------|------------|-------|
| LAS   | 60        | 10                | 15   | 24     | 11         | 1     |
| Plaid | 60        | 10                | 3    | 17     | 2          | 0     |
| CC    | 60        | 8                 | 0    | 12     | 2          | 0     |
| ISA   | 72        | 2                 | 18   | 23     | 5          | 0     |
| OPSM  | 15        | 0                 | 0    | 3      | 0          | 0     |
| SAMBA | 289       | 15                | 20   | 72     | 10         | 1     |
| BiMax | 1977      | 329               | 0    | 0      | 0          | 0     |

and Genomes, Kanehisa and Goto (2000), http://www.genome.jp/kegg/]. For the latter we considered cytobands, which consist of disjoint groups of genes such that the genes in a group have contiguous genomic locations. Definitions of KEGG and cytoband categories were taken from R metadata packages on Bioconductor (Bioconductor v 1.9, packages hgug4110b and hgu95av2).

For each bicluster gene set we computed a Bonferroni corrected hypergeometric $p$-value to assess its overlap with each KEGG category, and computed a similar $p$-value for each cytoband. We considered 153 KEGG categories and 348 cytobands that contained at least 10 genes (post filtering) on our sample arrays. A gene set was deemed to have significant overlap if any of the p-values computed in this way was less than $10^{-10}$. This threshold was selected to adjust for the anti-conservative behavior of the hypergeometric test in the presence of even moderate levels of gene correlation [see Barry, Nobel and Wright (2005) for more details]. The column *Gene* of Table 3 shows the number of biclusters having signficant overlap with at least one KEGG category or cytoband.

The third test concerns the differential expression of KEGG and cytoband categories across the sample set of a bicluster and its complement. From each bicluster we formed a treatment group consisting of the samples in the bicluster and a control group consisting of the complementary samples that are not in the bicluster. We tested for KEGG categories showing differential expression across the defined treatment and control groups using the SAFE procedure of Barry, Nobel and Wright (2005), and counted the number of categories passing the test at the 5% level. The permutation based approach in SAFE accounts for multiple comparisons and the (unknown) correlation among genes. A similar testing procedure was carried out for cytobands.

If no KEGG category were differentially expressed across the treatment and control groups corresponding to a particular bicluster, roughly 5% of the categories would exhibit significant differential expression by chance. We considered a bicluster sample set to yield significant differential expression of KEGG categories if the number of significant categories identified by SAFE exceeds the 5th percentile of the Bin(153, 0.05) distribution. An analogous determination was made for cytobands. The number of biclusters whose sample sets yield significant differential expression for KEGG categories or cytobands is presented in the *Sample* column of the Table 3.

The rightmost columns of Table 3 show the number of biclusters passing two or three tests. From an exploratory point of view, these biclusters are of potential interest, and represent a natural starting point for further experimental analysis. Accounting for the number (or effective number) of biclusters produced by each method, specifically the large output of SAMBA and the small output of OPSM, LAS outperformed the other methods under study, particularly in regards to biclusters satisfying two out of the three tests.

4.5. *Classification.* Biclustering algorithms identify distinguished sample-variable associations, and in doing so, can capture useful information about the data. To assess how much information about disease subtypes and ER status is captured by the *set* of biclusters produced by different methods, we examined the classification of disease subtypes using patterns of bicluster membership in place of the original expression measurements. Similar applications of biclustering for the purpose of classification were previously investigated in Tagkopoulos, Slavov and Kung (2005) and unpublished work [Grothaus (2005); Asgarian and Greiner (2006)].

Once biclusters have been produced from the data matrix, we replaced each sample by a binary vector whose $j$th entry is 1 if the sample belongs to the $j$th bicluster, and 0 otherwise. A simple $k$-nearest neighbor classification scheme based on weighted Hamming distance was applied to the resulting binary matrix: the classification scheme used the subtype assignments of training samples to classify unlabeled test samples. The number of rows in the derived binary matrix is equal to the number of biclusters; in every case this is far fewer than the number of genes in the original data.

To be more precise, let $X = [x_1, \ldots, x_n]$ be an $m \times n$ data matrix, and let $C_1, \ldots, C_K$ be the (index sets of) the biclusters produced from $X$ by a given biclustering method. We map each sample (column) $x_i$ into a binary vector $\pi(x_i) = (\pi_1(x_i), \ldots, \pi_K(x_i))^t$ that encodes its bicluster membership:

$$\pi_k(x_i) = \begin{cases} 1, & \text{if } x_i \text{ belongs to the sample set of bicluster } C_k, \\ 0, & \text{otherwise.} \end{cases}$$

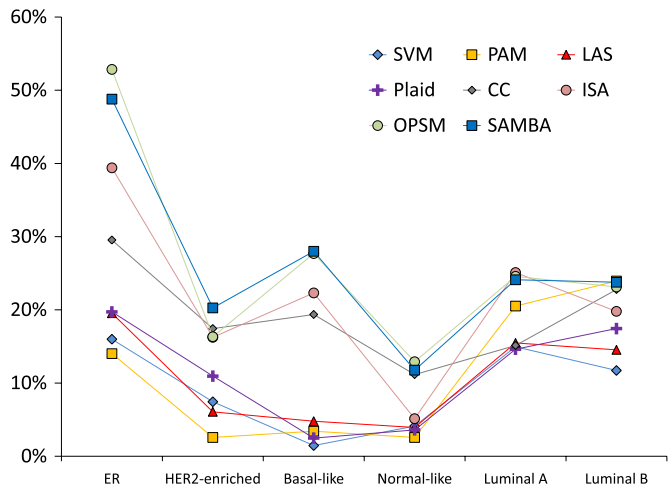

Fig. 5. *Classification error rates for SVM on the original data and the 5-nearest neighbor with weighted Euclidean distance applied to the "pattern" matrix.*

The original data matrix $X$ is then replaced by the $K \times n$ "pattern" matrix $\Pi = \{\pi(x_1), \ldots, \pi(x_n)\}$. In the Hu data, for example, the 13,666 real variables in $X$ are replaced by $K < 300$ binary variables in $\Pi$. Subtype and ER designations for the initial data matrix $X$ carry over to the columns of $\Pi$.

For each of the breast cancer subtypes in the Hu data, we used 10-fold cross validation to assess the performance of a 5-nearest neighbor classification scheme applied to the columns of the binary pattern matrix $\Pi$. The nearest neighbor scheme used a weighted Hamming distance measure, in which the weight of each row is equal to the square of the $t$-statistic for the correlation $r$ between the row and the response, $t^2 = (n-2)r^2/(1-r^2)$. In each case, the weights were calculated using only the set of training samples. For each subtype, the average number of cross-validated errors was divided by the total number of samples, in order to obtain an overall error rate. The results are displayed in Figure 5. For comparison, we include 10-fold cross validation error rates of a support vector machine (SVM) classifier applied to the original expression matrix $X$. As the figure shows, the error rates of LAS and Plaid are similar to those of SVM across the phenotypes under consideration. Using the pattern information from 60 biclusters, LAS and Plaid were able to distinguish individual subtypes with the same degree of accuracy as SVM applied to the original data with 13,666 variables.

4.6. *Lung data.* Validation results for the lung cancer data are contained in the Supplementary Materials [Shabalin et al. (2009)]. They are similar to the results for the breast cancer data considered above. The principal difference was the improved performance of ISA in tests of subtype capture.

While ISA biclusters continued to have small sample sets, the disease subtypes for the lung data contained fewer samples than those in the breast data.

**5. Simulations.** In addition to real data, we also investigated the behavior of the LAS algorithm on a variety of artificially created datasets. Our primary goals were to assess (i) the ability of the algorithm to discover significant submatrices under the additive model (1), (ii) the stability of the algorithm with respect to the initial random number seed, and (iii) the sensitivity of the algorithm to noise. The results of these simulations are described below. All relevant figures and tables (except Table 4) appear in the Supplementary Materials [Shabalin et al. (2009)].

5.1. *Null model with one embedded submatrix.* The key step of the LAS algorithm is to identify a submatrix of a given matrix that maximizes the score function. The approach taken by LAS is heuristic. As there are no efficient algorithms for finding optimal matrices, outside of small examples, we cannot check directly if the submatrix identified by LAS is optimal. In order to evaluate the LAS search procedure, we generated a number of data matrices of the same size as the Hu dataset, with i.i.d. $N(0,1)$ entries. For $k = 4, 8, 16, \ldots, 4096$ and $l = 4, 8, 16, 32$, we added a constant $\alpha = 0.1, 0.2, \ldots, 1$ to a $k \times l$ submatrix of the initial Gaussian matrix. The basic LAS search was carried out on each of the $11 \times 4 \times 10 = 440$ resulting matrices, and was considered a success if the search returned a bicluster whose score was at least as high as that of the embedded submatrix. The LAS search failed in only three cases; in each the embedded submatrix had relatively low scores (less than 100, while scores of other submatrices ranged up to 61,415.5). The search procedure was successful in all cases when the number of iterations used in the procedure was increased from 1000 to 10,000.

5.2. *Null model with multiple embedded submatrices.* We also tested the ability of LAS to discover multiple embedded submatrices. Simulations were performed with a varying number of embedded biclusters (up to 50), with 10 simulations for each number of biclusters. In each simulation, we first generated a $1000 \times 1000$ Gaussian random matrix. Then we randomly selected size and position of each bicluster, independently assigning rows and columns of the matrix to the bicluster with probability 0.02, so that the expected size of a bicluster is $20 \times 20$. Biclusters were generated independently, allowing for overlap. The elements of every bicluster were raised by 2 and LAS was applied to the resulting matrix. LAS was set to search for the correct number of biclusters with the default 1000 iterations per bicluster.

Table 4
*Discovery of multiple biclusters*

| Number of biclusters | Average match |
|---|---|
| 1 | 1.000 |
| 2 | 0.997 |
| 3 | 0.997 |
| 4 | 1.000 |
| 5 | 0.998 |
| 10 | 1.000 |
| 15 | 0.999 |
| 20 | 0.999 |
| 30 | 0.993 |
| 50 | 0.989 |

For every embedded bicluster $U$, we assessed its overlap with each detected bicluster $\tilde{U}$ using the minimum of specificity $|U \cap \tilde{U}|/|U|$ and sensitivity $|U \cap \tilde{U}|/|\tilde{U}|$, equivalently, $|U \cap \tilde{U}|/\max(|U|,|\tilde{U}|)$, and matched $U$ with the closest $\tilde{U}$. The average overlap across embedded biclusters and simulations for each number of true biclusters is presented in Table 4. The numbers indicate consistent accuracy of LAS in the detection of multiple embedded biclusters.

5.3. *Stability.* In order to check the stability of LAS with respect to the randomization used in the basic search procedure, we ran the algorithm 10 times on the Hu dataset with different random seeds. In order to assess the performance of the algorithm, rather than its raw output, for each of the 10 runs we calculated the validation measures from Section 4: effective number of biclusters, average size, $p$-values for subtype capture (as in Section 4.3.1), and the number of biclusters that passed different biological tests (as in Section 4.4). The results are presented in the Supplementary Materials [Shabalin et al. (2009)]. There is little variation in the calculated measures across different runs of the algorithm. The effective number of biclusters ranged from 48.2 to 49.0, and average size ranged from $355 \times 26$ to $363 \times 27$. The number of biclusters with significant survival ranged from 9 to 13, and the number of biclusters having significant overlap with at least one KEGG category or cytoband ranged from 13 to 16. The SAFE analysis is computationally intensive, so we did not perform it for these simulations. Although the output of LAS is not deterministic, its summary statistics for average size and overlap are stable, and it is consistently successful in capturing cancer subtypes.

5.4. *Noise sensitivity.* In order to assess the effects of noise on the LAS output, we added zero mean Gaussian noise with standard deviation $\sigma = 0, 0.1, 0.2, \ldots, 1$ to the normalized Hu dataset (after tail transformation and column standardization). The resulting matrix was then column standardized, and LAS was applied to produce 60 biclusters.

For each level of noise we calculated validation measures for the LAS output; these are presented in the Supplementary Materials [Shabalin et al. (2009)]. As the level of noise increases, the average number of genes in the LAS biclusters decreased, as did the number of biclusters with having a significant association with Cytoband or KEGG categories. However, within the tested range of noise levels, the average number of samples did not change noticeably, and the subtype capture performance did not markedly decrease. The results indicate both high noise resistance of LAS and the strength of subtype signal.

**6. Discussion.** Biclustering methods are a potentially useful means of identifying sample-variable associations in high-dimensional data, and offer several advantages over independent row–column clustering. Here we have presented a statistically motivated biclustering algorithm called LAS that searches for large average submatrices. The algorithm is driven by a simple significance-based score function that is derived from a Bonferroni corrected *p*-value under a Gaussian random matrix null model. We show that maximizing the LAS score function is closely related to minimizing the overall description length of the data in an additive submatrix Gaussian model.

The LAS algorithm operates in a sequential-residual fashion; at each stage the search for a submatrix with maximum score is carried out by a randomly initialized iterative search procedure that is reminiscent of EM type algorithms. The only operational parameters of LAS are the number of biclusters it produces before halting, and the number of randomized searches carried out in identifying a bicluster. In our experiments on real data, we found that 1000 randomized searches per bicluster were sufficient to ensure stable performance of the algorithm.

We evaluated LAS and a number of competing biclustering methods using a variety of quantitative and biological validation measures. On the quantitative side, LAS produced biclusters exhibiting a wide range of gene and sample sizes, and low to moderate overlap. The former feature implies that LAS is capable of capturing sample-variable associations across a range of different scales, while the latter indicates that distinct LAS biclusters tend to capture distinct associations. Other methods varied considerably in their sizes and overlap, with a number of methods producing biclusters having a small number of samples and genes.

Many LAS biclusters had significantly higher scores than biclusters obtained by more traditional approaches based on *k*-means and hierarchical clustering. This suggests that the constraints associated with independent row–column clustering (considering rows and columns separately, assigning each row or column to a single cluster) substantially limit the ability of these methods to identify significant biclusters, and that more flexible methods may yield substantially better results.

In regards to capturing disease subtypes, LAS was competitive with, and often better than, KM and HC. Other methods did not perform particularly well, though we note that ISA did a good job of capturing and classifying the smaller disease subtypes present in the lung cancer data. In tests for survival, over-representation of functional categories, and differential expression of functional categories, LAS outperformed the other biclustering methods. These tests, unlike the quantitative measures of size and overlap, were based on clinical and biological information.

The classification study in Section 4.5 shows that simple binary profiles of bicluster membership can contain substantive information about sample biology. In particular, nearest neighbor classification of disease subtypes using membership profiles derived from LAS or Plaid biclusters was competitive with a support vector machine classifier applied to the full set of expression data. We note that the biclustering methods applied here are unsupervised, and depend only on the expression matrix: none makes use of auxiliary information about samples or variables.

Our simulation study shows that the LAS search procedure is effective at capturing embedded submatrices (or more significant ones) having moderate scores. Although the search procedure makes use of random starting values, its performance is stable across different random seeds. The ability of the algorithm to capture subtypes does not substantially deteriorate when a moderate amount of noise is added to the data matrix.

While the validation of biclustering here has focused on gene expression measurements, it is important to note that LAS and other biclustering methods are applicable to a wide variety of high-dimensional data. In preliminary experiments on high density array CGH data produced on the Agilent 244k Human Genome CGH platform, LAS was able to capture known regions of duplications and deletion (data not shown). The dataset contained roughly 250 samples and 240,000 markers. We note that among the seven biclustering methods compared in the paper only the current implementations of LAS and Plaid were able to handle datasets of this size.

LAS biclusters capture features of the data that are of potential clinical and biological relevance. Although some findings, such as disease subtypes, are already known, very often the methods used to establish them involve a good deal of subjective intervention by biologists or disciplinary scientists. LAS provides a statistically principled alternative, in which intervention (such as selecting biclusters of potential interest) can take place after the initial discovery process is complete.

We note that the LAS score function and search procedure can readily be extended to higher dimensional arrays, *for example*, three-dimensional data matrices of the form $\{x_{i,j,k} : i \in [m], j \in [n], k \in [p]\}$. Related extensions of the Plaid model have been developed by Turner et al. (2005).

As noted in Section 1.1, our use of the large average criterion is motivated by current biological practice in the analysis of gene expression and related data types. The validation experiments in the paper establish the efficacy of the large average criterion, and the LAS search procedure, for standard gene expression studies, and there is additional evidence to suggest that the criterion will be effective in the analysis of CGH data. Nevertheless, we note that the large average criterion is one of many that may be used in the exploratory analysis of high dimensional data. Other criteria and methods can offer additional insights into a given data set of interest, and may provide valuable information in cases, and for questions, where the large average criterion is not appropriate.

## APPENDIX

**Minimum description length connection.** Let the data matrix $X$ be standardized, so that its elements have zero mean and unit variance, and let $U$ be the selected bicluster. The code describing the data matrix must describe both the bicluster (size, location, average of its elements) and the residual matrix.

It is not possible to code real-valued data precisely with a finite-length code, so we construct a code describing the data with a precision of $C$ binary digits after the decimal point.

The size of the submatrix $U$ is described by variables $k \in [m]$ and $l \in [n]$, so coding these variables requires $\log_2(mn)$ bits. (We ignore rounding issues here and in what follows.) There are a total of $\binom{m}{k}\binom{n}{l}$ different $k \times l$ submatrices in a $m \times n$ matrix, so the code describing the location of the submatrix requires $\log_2[\binom{m}{k}\binom{n}{l}]$ bits. To code the submatrix average $\tau$, we assume that it lies within the interval $[-8, 8]$ (we did not observe $|\tau| > 1.5$ in our experiments). Then the code describing the average $\tau$ of the submatrix $U$ takes $\log_2 16 + C = 4 + C$.

Finally, we describe the residual matrix. The data set is standardized, so its total variation (sum of squares) is $nm$. A $k \times l$ submatrix with average $\tau$ explains variation $\tau^2 kl$, so that the variation of the residual matrix is $nm - \tau^2 kl = nm[1 - \frac{kl\tau^2}{nm}]$. Thus, under model (1) the elements of the residual matrix are approximately distributed as $N(0, 1 - \frac{kl\tau^2}{nm})$.

Coding of a random variable $X$ with density $f(x)$ requires $-\log_2(f(X)) + C$ bits, or, on average, $-\int f(x) \log_2(f(x))\, dx + C$ bits. Let $C_N$ be the average code length for standard normal random variable, $C_N = -E \log_2(\phi(Z)) + C$, where $z \sim N(0, 1)$ and $\phi(z)$ is density of standard normal distribution. Then for $X \sim N(0, \sigma^2)$ the average code length is $C_N - \log_2(\sigma^2)/2$. Thus, coding of the residual matrix takes $nm[C_N - \log_2[1 - \frac{kl\tau^2}{nm}]]$ bits on average.

Combining the codelengths above, the length of the code describing the $X$ using a $k \times l$ bicluster $U$ with average $\tau$ is

$$MDL(U) = \log_2(nm) + 4 + C + \log_2\left[\binom{m}{k}\binom{n}{l}\right]$$

$$+ nm\left[C_N - \log_2\left[1 - \frac{kl\tau^2}{nm}\right]/2\right].$$

In the applications we considered, the explained variation was a small fraction (typically less than 1/1000) of the total variation. Thus, we can apply first order approximation: $\log_2[1 + x] = x/\ln(2) + o(x)$. Then

$$MDL(U) \approx const + \log_2\left[\binom{m}{k}\binom{n}{l}\right] - kl\tau^2/2\ln(2).$$

We pulled the constant terms and terms depending on $n$ and $m$ out, as they do not depend on the selected bicluster.

Let's now consider the LAS score function,

$$S(U) = -\log\binom{m}{k} - \log\binom{n}{l} - \log(\Phi(-\sqrt{v^2 kl})).$$

For large $x$ we can approximate $\Phi(-x) = \exp[-x^2/2]/x + o(x)$, getting

$$S(U) \approx \ln(2)\left[-\log_2\binom{m}{k}\binom{n}{l} + \tau^2 kl/2\ln(2) - \log_2(\tau^2 kl)/2\right].$$

Easy to see that except for the small factor of $\log_2(\tau^2 kl)/2$ the code length and score function approximations are linearly dependent:

$$S(U) \approx const - \ln(2) MDL(U).$$

**Running configurations for other methods.** All biclustering methods described in this paper were run on the same machine: AMD64 FX2 2.8GHz, 4GB RAM, running Windows XP Professional (64 bit). The same imputed dataset as run through LAS was loaded into the other programs. If a method was written in Java, the "Xmx1024m" key was added to the command line for proper memory allocation. In all cases, we preferred to use the default running parameters as given by the software used to run the algorithms (*BicAT* for BiMax, CC, ISA, OPSM, and *Expander* for SAMBA).

*Running parameters.* *Plaid*, as it is scripting based, a script was written to iterate over the steps *findm*, accept, shuffle 60 times, to produce 60 biclusters. *Cheng-Church*: $seed = 13$, $\Delta = 0.1$, $\alpha = 1.2$, *NumberOutput* $= 30$, *ISA*: $seed = 13$, $t\_g = 2$, $t\_c = 2$, $StartingNum = 100$, *OPSM*: $PassedModels = 10$, $BiMax$: $Gene_{\min} = 10$, $Sample_{\min} = 5$, $SAMBA$: try covering all probes, $OptionFiles = valsp\_3ap$, $OverlapPrior = 0.1$, $ProbesToHash = 100$, $Memory_{\max} = 500$, $HashKernel_{\min} = 4$, $HashKernel_{\max} = 7$. The $OverlapPrior$ value ensures that for each new cluster generated, its elements were 90% unique to any previously discovered bicluster.

## SUPPLEMENTARY MATERIAL

**Supplement to Finding Large Average Submatrices in High-Dimensional Data** (DOI: 10.1214/09-AOAS239SUPP; .pdf). The supplementary article contains additional tables and figures used in validation of LAS. It includes bar plots similar to Figure 4 produced for each cancer subtype, a complete set of validation results for the lung cancer gene expression data set from Bhattacharjee et al. (2001), and tables with simulation results measuring stability and noise sensitivity of the LAS model and algorithm.

A. Shabalin
A. Nobel
Department of Statistics
 and Operations Research
University of North Carolina
 at Chapel Hill
Chapel Hill, North Carolina 27599
USA
E-mail: shabalin@email.unc.edu
        nobel@email.unc.edu

V. Weigman
Department of Biology
University of North Carolina
 at Chapel Hill
CB #3280, Coker Hall
Chapel Hill, North Carolina 27599
USA
E-mail: victor@med.unc.edu

C. Perou
Lineberger Comprehensive Cancer Center
University of North Carolina
 at Chapel Hill
450 West Dr.
Chapel Hill, North Carolina 27599
USA
E-mail: cperou@med.unc.edu